\documentclass[final]{custom}
\usepackage{graphicx}
\usepackage{rotating}
\usepackage{amsfonts, amsmath, amsthm, amssymb}
\usepackage{mathptmx}
\usepackage[numbers, square, sort]{natbib}
\usepackage[ruled,vlined]{algorithm2e}

\makeatletter

\begin{document}

\newcommand{\Ricochet}{\textsc{Ricochet}}
\newcommand{\fa}{\textsuperscript{$\star$}}

\title{Development of data processing and analysis pipeline for the \Ricochet{}  experiment.}
\author{%
    \name{J. Colas\fa{}\thanks{\fa{}\,Email: j.colas@ip2i.in2p3.fr}, J. Billard, S. Ferriol, J. Gascon and T. Salagnac on behalf of the \Ricochet{} collaboration.}%
    \affil{Univ Lyon, Universit\'e Lyon 1, CNRS/IN2P3, IP2I-Lyon, F-69622, Villeurbanne, France}%
}

\maketitle

\begin{abstract}

    Achieving a percentage-level precision measurement of the Coherent Elastic Neutrino Nucleus Scattering (CE$\nu$NS) spectrum requires a robust data processing pipeline which can be characterised with great precision. To fulfil this goal we present hereafter a new Python-based data processing pipeline specifically designed for temporal data analysis and pulse amplitude estimation. This pipeline features a data generator allowing to accurately simulate the expected data stream from the \Ricochet{}  experiment at the Institut Laue Langevin (ILL) nuclear reactor, including both background and CE$\nu$NS signals. This data generator is pivotal to fully understand and characterise the data processing overall efficiency, its reconstruction biases, and to properly optimise its configuration parameters. We show that thanks to this optimized data processing pipeline, the CryoCube detector array will be able to achieve a 70~eV energy threshold combined with electronic/nuclear recoil discrimination down to $\sim$100~eV, hence fulfilling the \Ricochet{} targeted performance.


\end{abstract}

\section{Motivations}

The \Ricochet{}  experiment is a reactor neutrino observatory which will host two complementary detectors inside a shielded cryostat installed 8.8 meters away from the 58~MW ILL research nuclear reactor~\cite{scott}. One of these detector, the CryoCube presented in \cite{thomas}, is an array of 27 high purity Ge sub-detector. Each 40~g sub-element will be instrumented with a NTD-Ge thermal sensor and multiple aluminium electrodes for heat and charge measurements, respectively. Overall, the CryoCube will provide at most 150 data channels. Each detector channel will be connected to a High Electron Mobility Transistor (HEMT) low-noise cold pre-amplifier described in~\cite{Juillard:2019njs}. \\
The data taking strategy is to perform continuous stream acquisition at a sampling rate between 10~kHz and up to 100~kHz, as required for an efficient anti-coincident muon veto tagging.
The data acquisition is planned to be running for months continuously such that we expect $\mathcal{O}(100)$ terabyte of raw data for one operating year at ILL.\\
The processing will be done using analysis window with a duration between 0.5 to 2~s. The typical value of 1~s is well suited to contain completely the heat pulse without including multiple pulses in the same window. The heat pulses are characterised by the sum of three exponentials, each describing a time response of the detector thermal system. They are much slower ($\mathcal{O}(10)~\mu$s) than the ionisation pulses which can be described by a Heavyside function because of the very fast ($\mathcal{O}(10)$~ns) charge collection to the electrodes.

\section{Simulating the CryoCube}
\subsection{Heat and ionisation pulses}
\label{sec:simulation}

We expect to measure two types of events : electronic recoils (ER) and nuclear recoils (NR) which differ from their ionisation yield $Q=E_\text{ion}/E_\text{recoil}$. We use the quenching model given by Lindhard for NR \cite{lindhard1963integral} and $Q=1$ for ER. A third type of event may occur in our detector, which are the so-called \textit{heat-only} (HO) event. These events form a low energy excess background and come from a yet unidentified source~\cite{billard:tel-03259707} and are characterised by a null quenching factor $Q=0$. When the electrodes are biased to a non-zero electric potential we have to consider the Neganov-Trofimov-Luke (NTL) effect which corresponds to an additional heat contribution produced by the drift of the charges in the Ge crystal under the electric field. With this additional contribution, the energy relations can be written as \cite{Armengaud_2017}
\begin{equation*}
    E_\text{ion} =  E_\text{recoil} \cdot  Q  \ \ \ \ \ \ \ \ \
    E_\text{heat}  =  E_\text{recoil} \cdot ( 1 +  Q  \frac{V}{\epsilon}  )
\end{equation*}

\begin{figure}[h!]
    \begin{center}
        \includegraphics[width=0.32\linewidth]{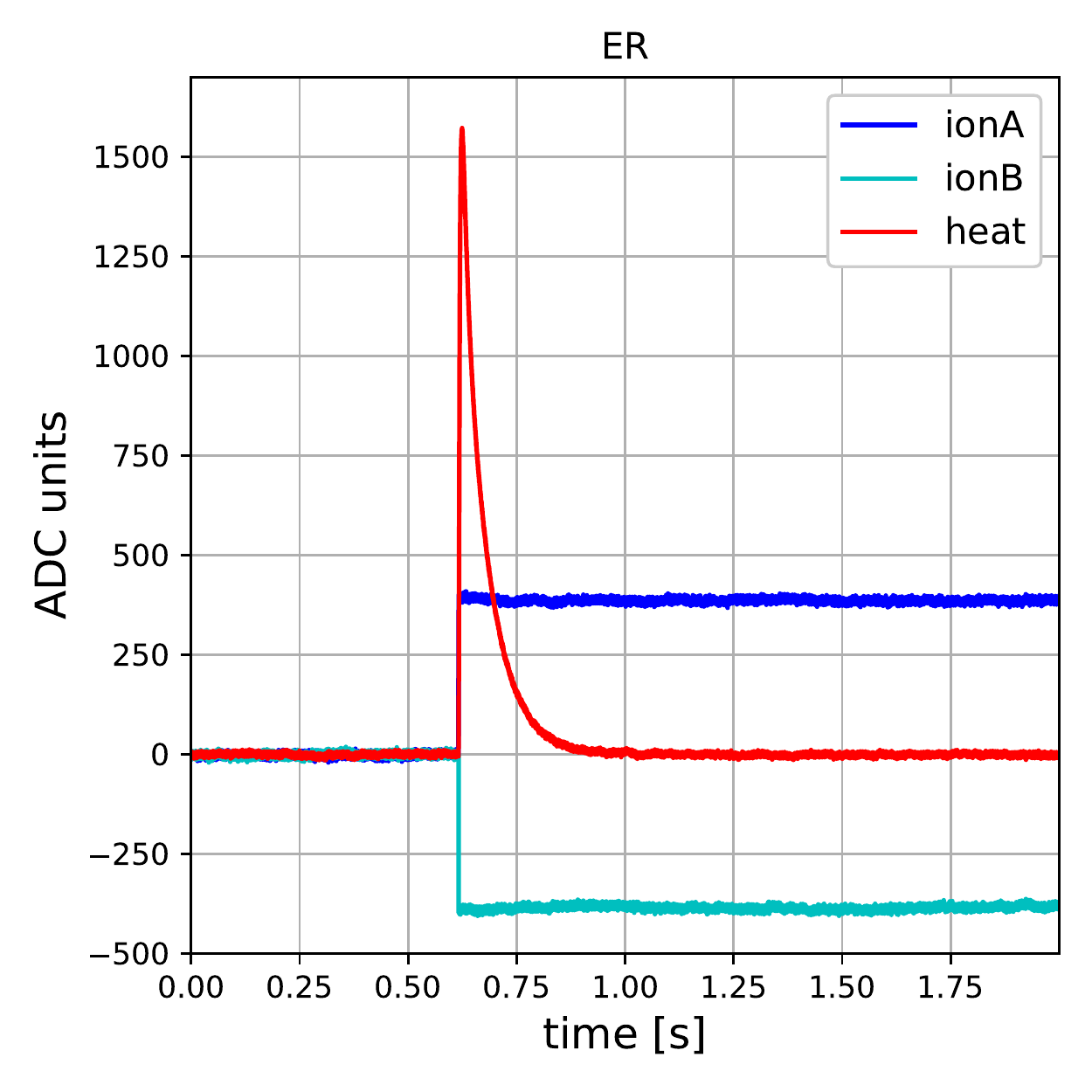}
        \includegraphics[width=0.32\linewidth]{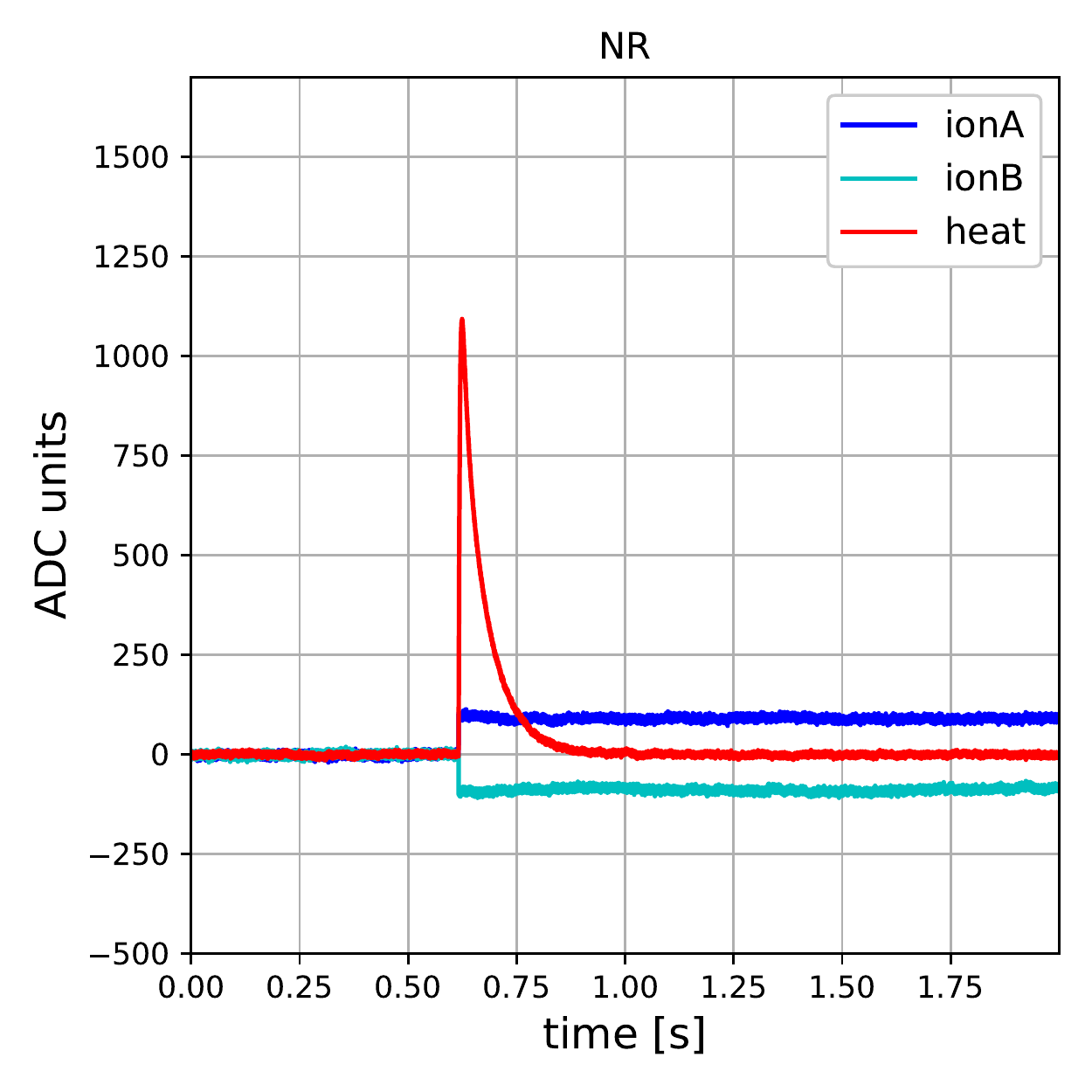}
        \includegraphics[width=0.32\linewidth]{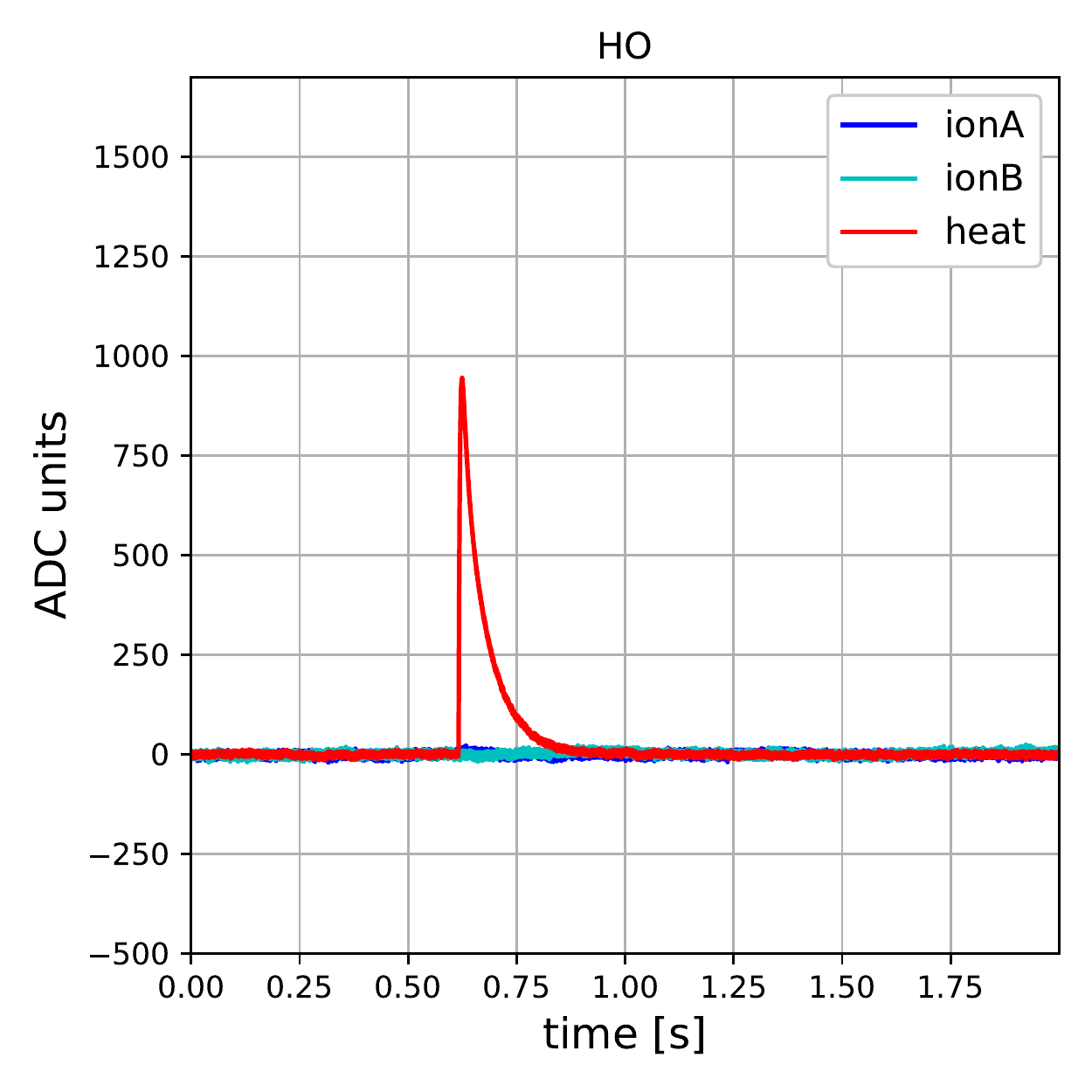}
    \end{center}
    \caption{Simulation of a 10~keV recoil energy event considering a Ge detector with a planar design and a 2V potential between electrodes : \textit{Left} - Electronic recoil. \textit{Center} - Nuclear recoil. \textit{Right} - "Heat Only" event.}
    \label{fig:pulse_simu}
\end{figure}

The simulation of an ER, NR and HO event of 10~keV recoil energy is shown in Fig. \ref{fig:pulse_simu} with $\epsilon=3$~eV (Ge) and $V=2$~V. The NTL effect is clearly visible since the amplitude of the heat pulse increases with the amplitude of the ionisation signals. The noise added to these traces corresponds to the noise expected for the future CryoCube electronic and is described in the following subsection.

\subsection{Electronic noise generation}

\begin{figure}[h!]
    \centering
    \includegraphics[width=0.6\linewidth]{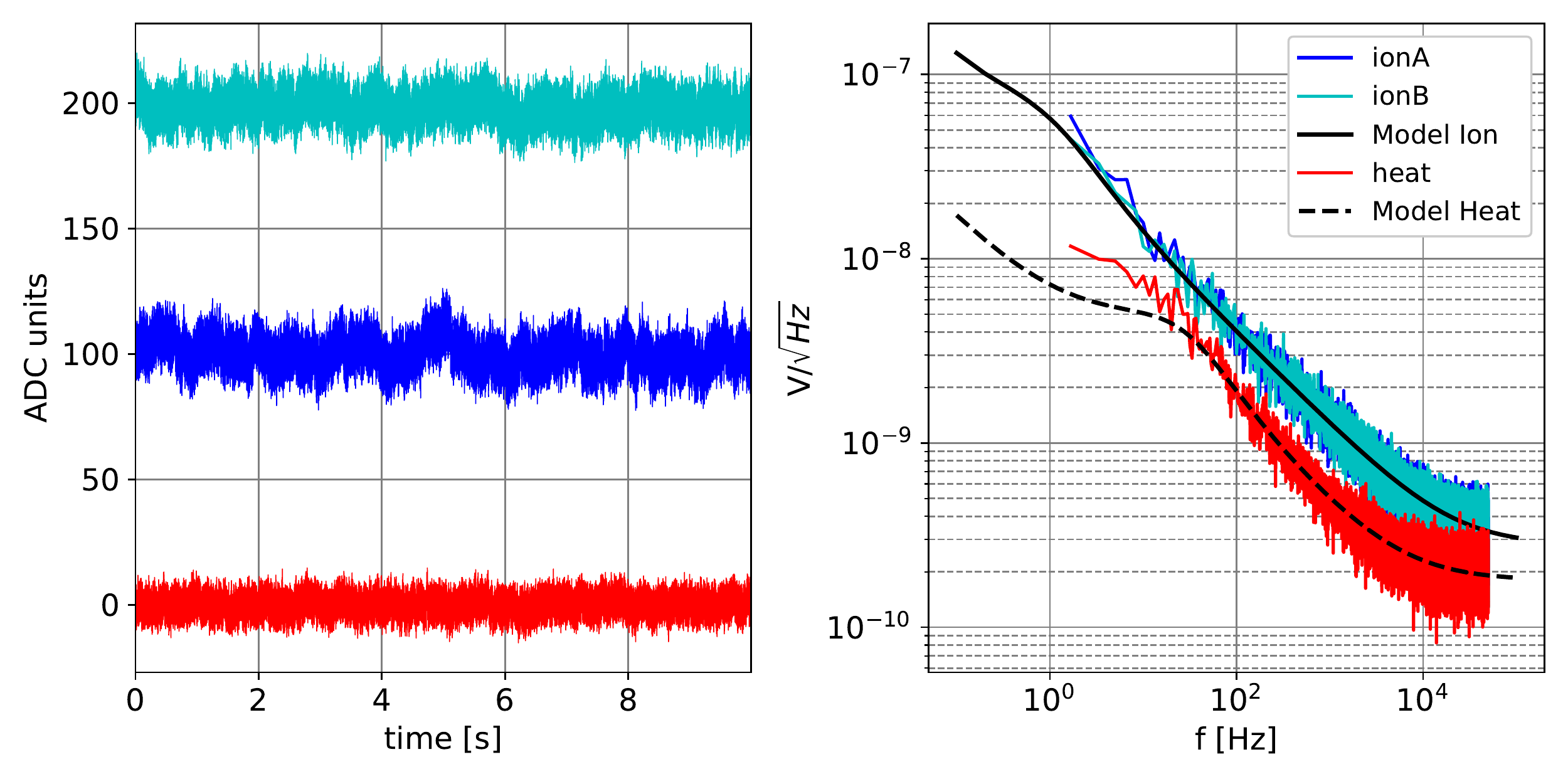}
    \caption{\textit{Left} Simulation of noises from our heat and ionisation models (see~\cite{Juillard:2019njs}) using 2 s data chunks and 10 s of total duration at 100 kHz. \textit{Right} Corresponding linear Power Spectral Density (LPSD). The LPSD is the square root of the PSD.}
    \label{fig:noises}
\end{figure}

The noise models for both the ionisation and heat channels are extracted from~\cite{Juillard:2019njs} and can be accurately reproduced from the following three noise structures: 1) {white noise}, generated using random normal distributed samples, PSD $\propto $cst; 2) {brown noise}, generated using first order low-pass filtered white noise, PSD $\propto f^{-2}$; and 3) {pink noise}, generated using the Voss-McCartney algorithm \cite{Voss19781fNI}, PSD $\propto f^{-1}$. These noises, characterised by their power spectral density (PSD, in V$^2$/Hz), will be used as our basis for our noise modeling such that the expected heat and ionisation noises can be approximated by:
\begin{align}
    \text{PSD}_\text{heat/ion}^\text{expected}(f) & \sim A_\text{heat/ion} + B_\text{heat/ion}/f + C_\text{heat/ion}/f^2
\end{align}

This approximation could be improved using more different noise models or additional filter, but this is already sufficient to reproduce very well the ionisation noise and relatively well the heat noise as we we can see on Fig.~\ref{fig:noises}. The discrepancy between model and generated noise around  f=10 Hz shows that a high-pass filter should be used to reduce the gap between model and generated noise. But since the implementation of such filter can be tricky for bigger-than-memory signals we will develop this feature for a future work. To be able to generate hours-long signal, we had to adapt our algorithms for chunk-by-chunk data generation where the data stored in memory at a time (a chunk) is not too large and fits in RAM. Our algorithms preserve the phase continuity between each consecutive chunks.

\section{\textbf{Processing data}}

\begin{figure}[h!]
    \centering
    \includegraphics[width=0.8\linewidth]{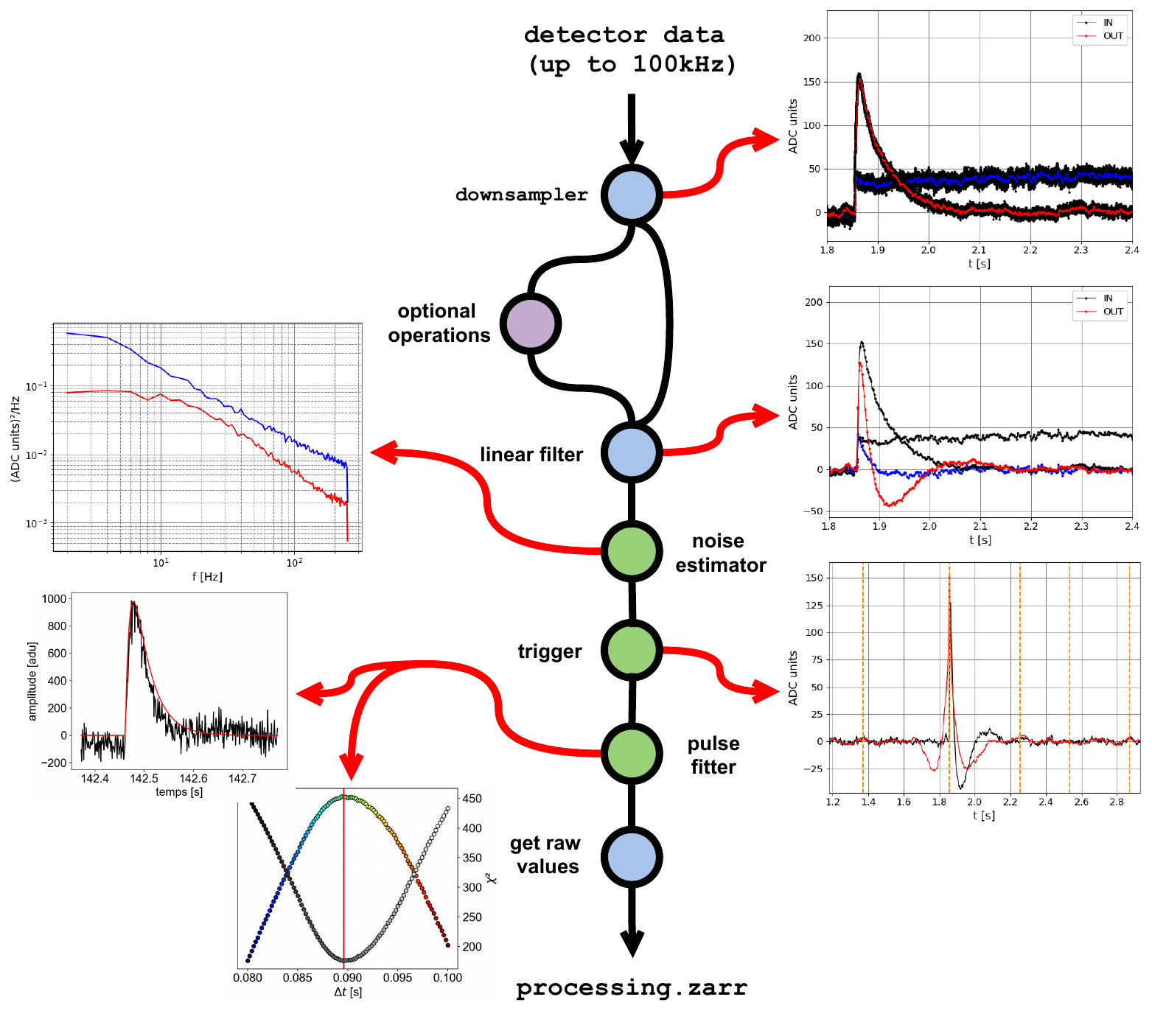}
    \caption{Schematic of the proposed data processing pipeline based on matched filtering for the \Ricochet{} 's CryoCube detector.}
    \label{fig:MPS_pipeline}
\end{figure}

The processing pipeline for the CryoCube was firstly characterised using simulated stream generated using the previously discussed approach. Using simulated realistic data is the best way to know efficiently the performances of such complex pipeline in terms of amplitude, timing resolution and trigger efficiency. The goal of our pipeline is to give an estimation of the amplitude $\hat{a}$, the location $\hat{t_0}$ and the quality of fit $\chi^2$ for each triggered pulses. The only required prior information is the analytical pulse shape for each channel. This knowledge gives us a starting point to evaluate the noise in the data stream which is the most important quantity for the estimation of $\hat{a}$ and $\hat{t_0}$ as we will explain in the following section. \\

The global view of the processing pipeline is presented in Fig. \ref{fig:MPS_pipeline}. Firstly the pipeline is setup using an external configuration yaml file. Then each hour of data is sent to this pipeline and undergoes: 1) a downsampling to reduce the sampling frequency $f_s$ (if needed for speed consideration), 2) a linear high-pass filtering to clean the low frequency end of the spectrum, 3) an estimation of the noise power spectrum, 4) a trigger event positions, and 5) an amplitude and position ($\hat{a}$, $\hat{t_0}$) estimator for each triggered events. The output and the input of this pipeline are compressed structured files in zarr format \cite{zarr_git}, which support multiple compression algorithm as well as simultaneous read and write. \\

\subsection{Noise estimation}
To estimate the noise PSD, we use the analytical pulse shape and look at the parts of the stream which least resemble to pulses. One way to do that is to use the correlation between a pulse trace with the entire stream and look for the minimum amplitude of the obtained signal. We then take the maximum number of traces which have no data point above a particular threshold dynamically defined. Typically in our case, this first step of the noise estimator returns at least 100 noise traces (for a one hour stream ??) from which a first PSD $J_0$ can be estimated. The second step is to take the maximum number of traces from the stream and compute the $\chi^2$ distribution such that $\chi^2_i = \sum_f \frac{\vert T_i(f) \vert^2}{J_0(f)}$ where $T_i$ is the i-th trace in the Fourier space normalised as a LPSD. This gives us a $\chi^2$ distribution with a peak around the number of frequency bins, corresponding to all the traces similar to the first ones considered for $J_0$. These are the \textit{good} noise traces from which the final noise PSD $J$ can be estimated with higher statistics, hence lower variance.

\subsection{Triggering potential events}

The trigger algorithm is based on matched filtering and is close to the implementation presented in the widely used reference \cite{di2011lowering}. The idea is to apply a matched filter (matched to the pulse template) which is characterised by a transfer function $H$ such as $H(f) = \alpha \frac{S^*(f)}{J(j)}$, where $S^*$ is the complex conjugate of the Fourier transform (FT) of the pulse trace and $\alpha$ is a complex normalisation constant :
\begin{equation}
    \alpha = \underbrace{\left( \sqrt{\frac{N}{2f_s}}\right)}_{\text{FT-to-PSD term}} \underbrace{\left( \sum \frac{\vert S \vert^{2}}{J}\right)^{-1}}_{\text{amplitude term}}  \underbrace{\exp \left(i2\pi i_\text{0} \frac{f}{f_s}\right)}_{\text{phase term}}
\end{equation}
The resulting signal $m(t) = \mathcal{F}^{-1}(H * \mathcal{F}(x)(f))$ of the matched filtering of the data stream $x(t)$ is then sorted by decreasing values. The first value of this sorted array is retained as trigger position $i_0$ and a region around this position is discarded. Typically the size of this region is the same as processing window $N$ such that there is never two triggers in the same window. Then we repeat this first step by taking the first non-discarded value and append it to $i_0$ until all the data are discarded. By doing the trigger selection this way, we do not need to specify a particular threshold value and we are able to trig on the lowest pulse amplitudes, maximising our detection efficiency to the lowest event energies.

\subsection{Estimation of $\hat{a}$ and $\hat{t_0}$}

The trigger algorithm returns an array of integer $\Vec{i_0}$ representing the position in sample numbering where an event might be. We will use this information as a starting point to our fitting approach. The idea is to use the analytical pulse shape to create a 2-dimensional matrix $M_{i,k}$ with $i$ the index of starting $\tau_i$ (more explanations below) and $k$ is the window sample index between 0 and $N-1$. In other words this matrix contains the traces of the pulses for each channels and for different starting time $\tau_i$. This means that we evaluated the pulses functions on a grid of starting time $\tau_i$ thin enough to reach sub sample resolution. Then, to find the optimal starting time $\hat{t_0}=\tau_i\vert_{d\chi^2/dt=0}$ and amplitude $\hat{a}$ we minimise the $\chi^2$ defined, for one channel, as :
\begin{equation}
    \chi^2(\tau_i, \hat{a}) = \sum_f \frac{\vert D(f) - \hat{a} \mathcal{M}(\tau_i,f)) \vert^2}{J(f)}
\end{equation}
where $D$ is the FT of one window returned by the trigger normalised as LPSD, like $\mathcal{M}$ which is the LPSD normalised FT of the template matrix $M$. The minimisation is performed by calculating the optimal $\hat{a}$ for each $\tau_i$ and then we take the minimal $\chi^2$ value computed over the whole time range. Lastly, this simple example can be generalized to more than one channel, as done in the following, in building a global $\chi^2$ with its corresponding matrix of PSDs also including the cross-PSD (XPSD) between channels. 

\section{\textbf{Results}}
\subsection{Noise estimation and trigger efficiency}

\begin{figure}[h!]
    \centering
    \includegraphics[width=0.42\linewidth]{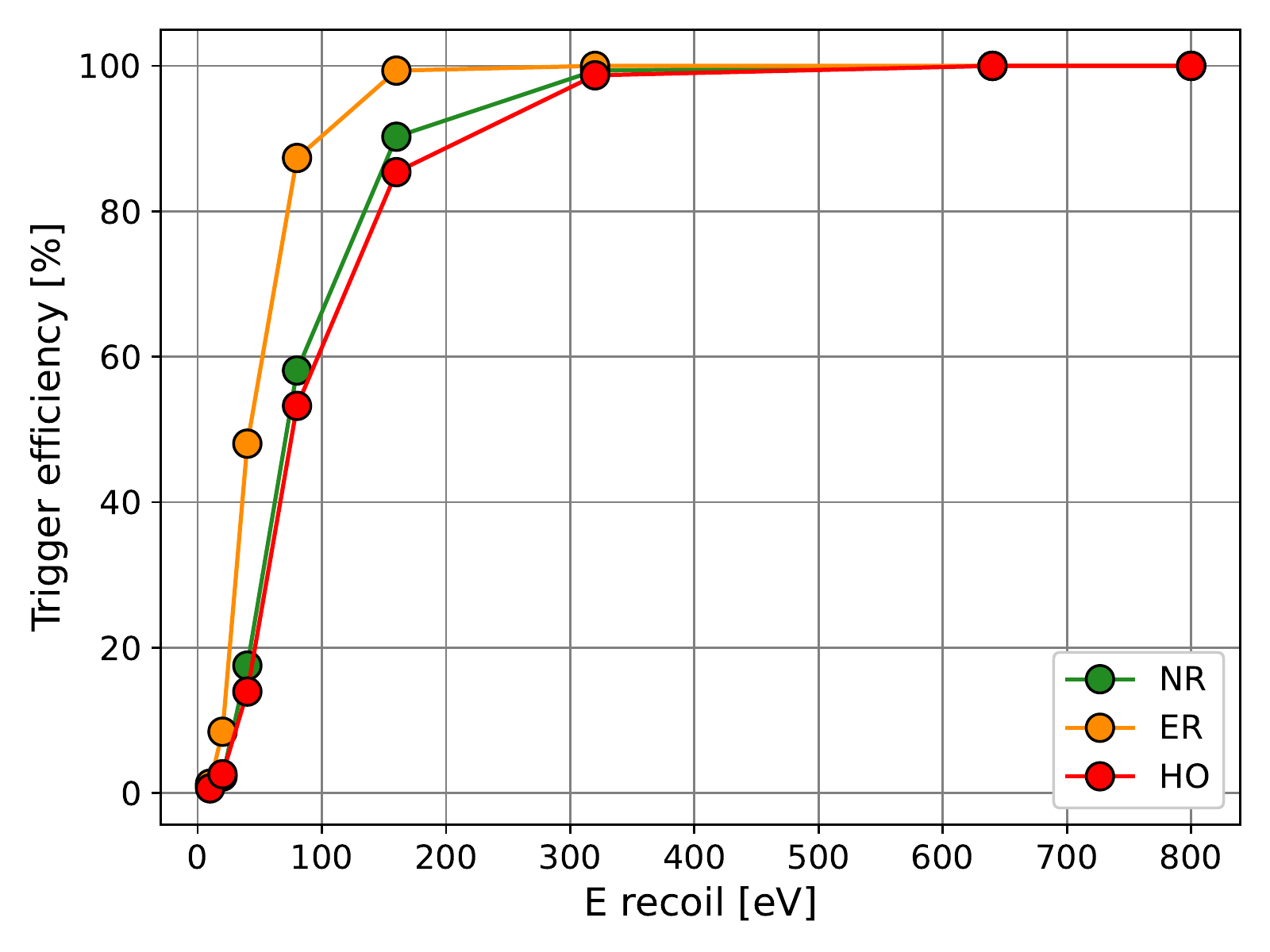}
    \includegraphics[width=0.42\linewidth]{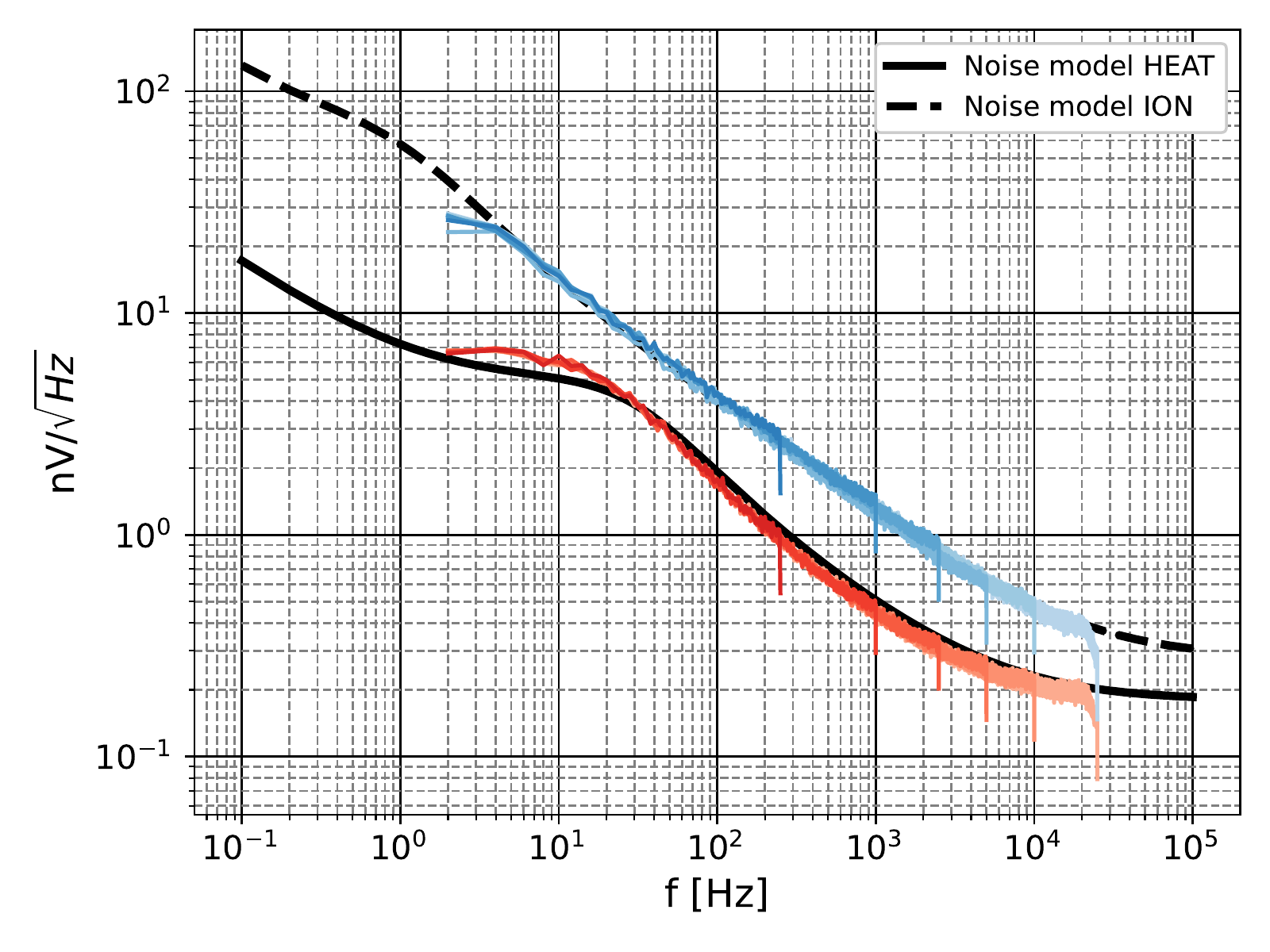}
    \caption{\textit{Left}- Trigger efficiency at $f_s = 1$kHz for different recoil energy based on the noise prediction \cite{Juillard:2019njs}. \textit{Right}- Estimated linear noise power spectrum density for heat (red) and ionisation channel (blue) using our proposed minimal correlation algorithm for identification of noise, at multiple sampling frequencies and compared to the noise models we tried to reproduce (see Fig. \ref{fig:noises}).}
    \label{fig:efficiencies}
\end{figure}

Applying the processing pipeline described above to simulated yet realistic data as presented in Sec.~\ref{sec:simulation} allows us to estimate the performance of the future CryoCube detector with the present processing scheme. We find a 50\% efficiency at about 70 eV nuclear recoil energy, with a slightly lower threshold for ER thanks to their higher NTL heat energy boost (see Fig.~\ref{fig:efficiencies}). We also see, from Fig.~\ref{fig:efficiencies}, that the noise estimation algorithm recovers properly the injected noise power spectral densities for all down-sampled sampling frequencies.

\subsection{Energy and Timing resolution}

\begin{figure}[h!]
    \centering
    \includegraphics[width=0.42\linewidth]{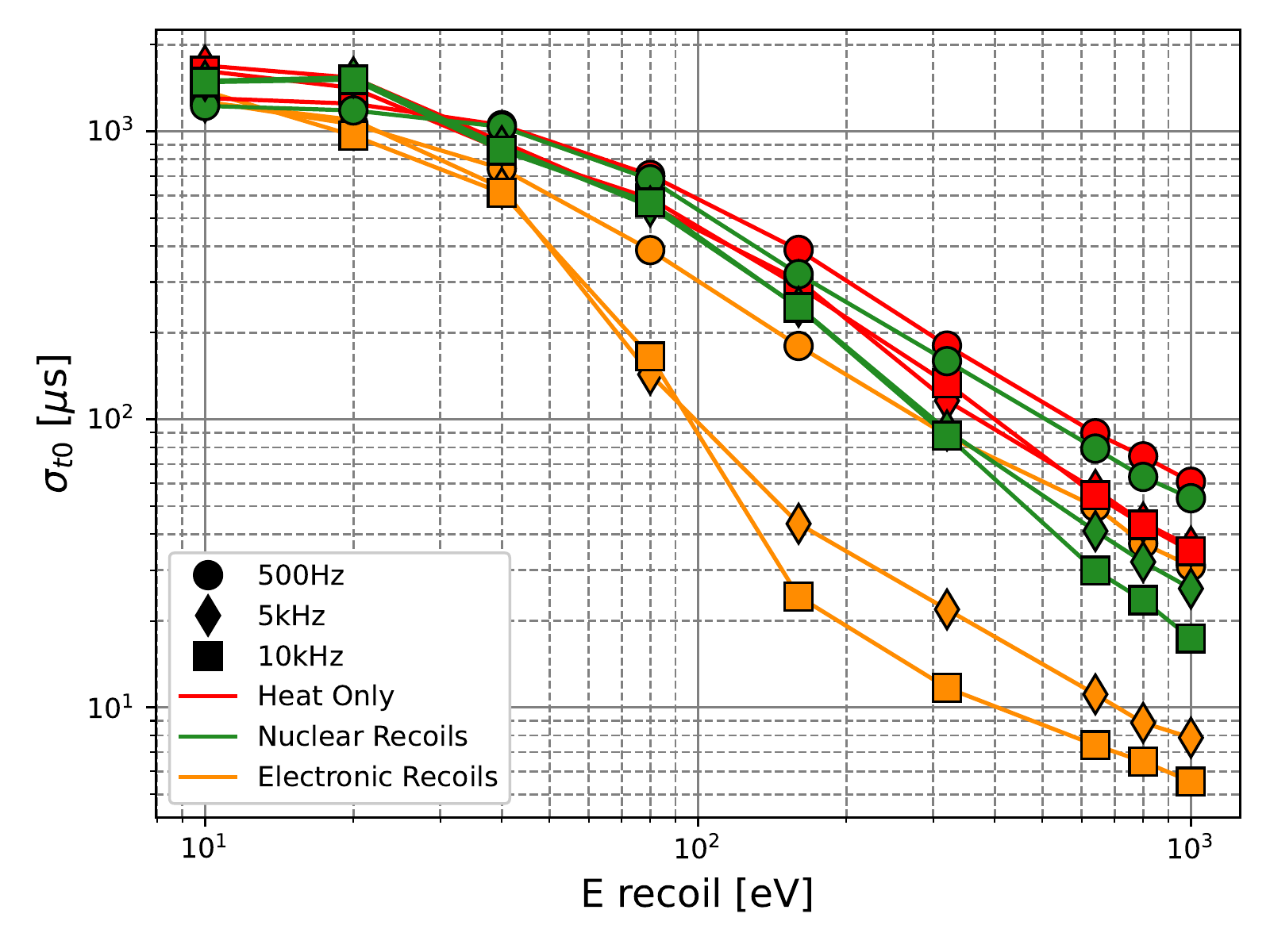}
    \includegraphics[width=0.42\linewidth]{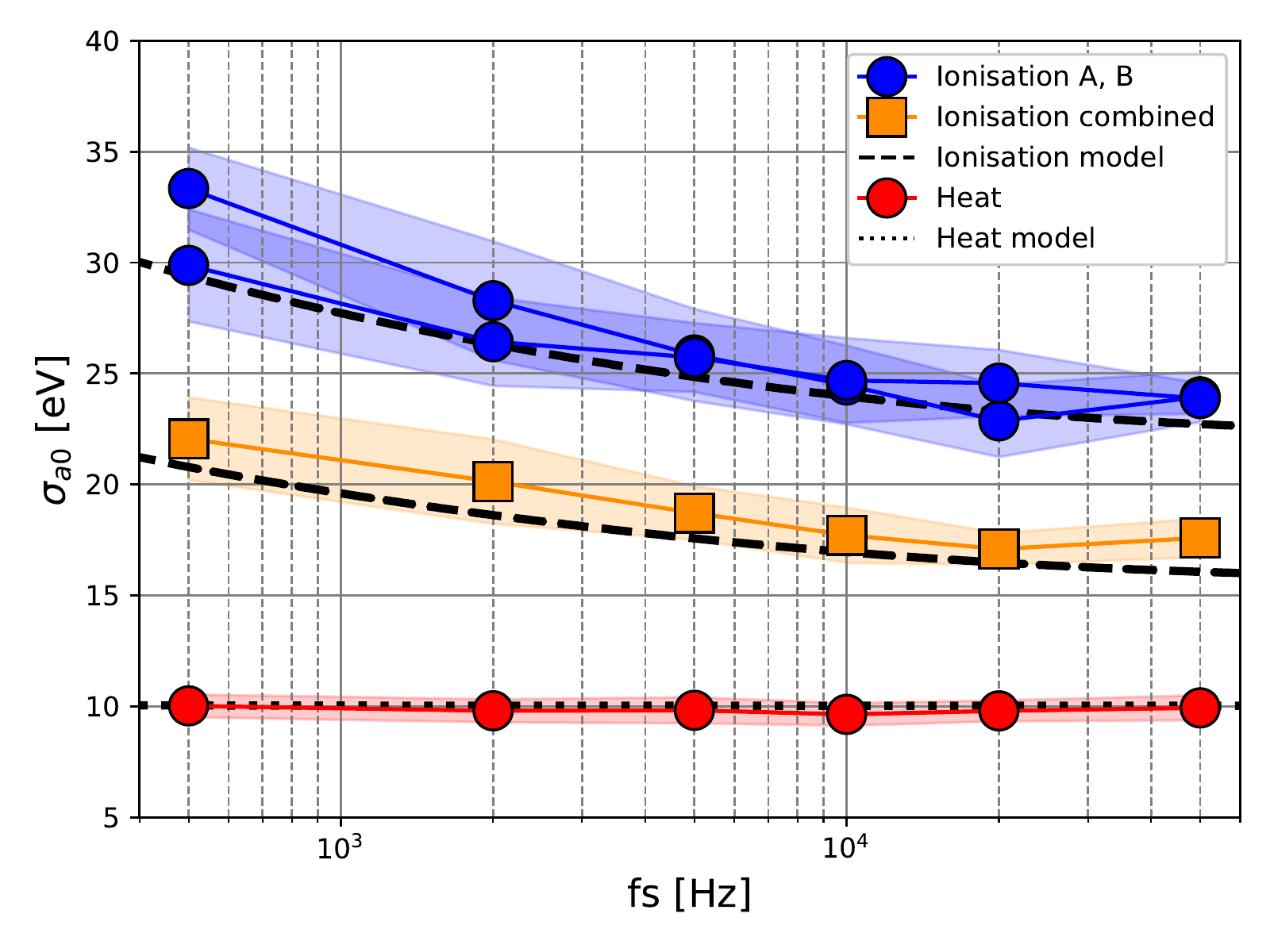}
    \caption{\textit{Left}- Timing resolution for different type of events versus the recoil energy. \textit{Right}- Amplitude resolution for heat and ionisation channels versus the sampling frequency.}
    \label{fig:resolutions}
\end{figure}

Fig.~\ref{fig:resolutions} (left panel) shows the timing resolution $\sigma_{t_0}$ as a function of the recoil energy, for varying sampling frequencies and recoil types. As one can see, $\sigma_{t_0}$ improves faster than $1/E_\text{recoil}$ thanks to the additional fast ionisation signal. This is particularly true for ER events because of their higher ionisation yield. Our expected timing resolution suggests that a 400~Hz muon veto trigger rate, as expected with \Ricochet{}  at ILL, is manageable.
From Fig.~\ref{fig:resolutions} (right panel), we can see that the baseline ionisation energy resolution $\sigma_{a_0}$ slightly improves with increasing the sampling frequency $f_s$  as expected from its higher signal bandwidth. The estimated baseline resolution is compatible with our theoretical calculations : $\sim$ 10 eV (heat), $\sim$ 20 eVee (ionisation) \cite{Juillard:2019njs}. As a result, this study confirmed the choice of 10~kHz as an optimal sampling frequency taking into account the quantity of raw data as well as the timing and energy resolutions~\cite{billard:tel-03259707}.

\subsection{Particle Identification capabilities}

\begin{figure}[h!]
    \centering
    \includegraphics[width=0.8\linewidth]{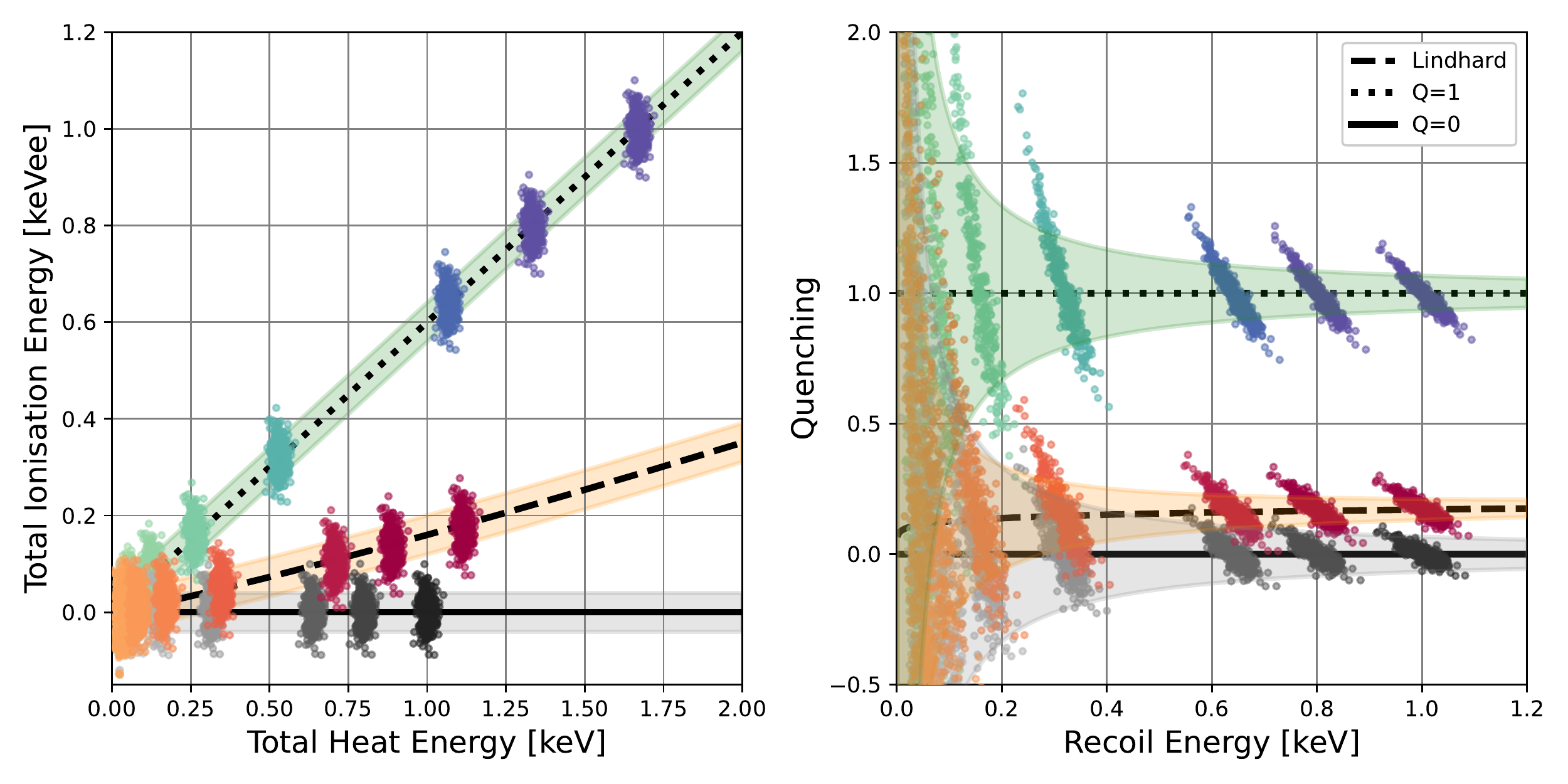}
    \caption{Quenching performances obtained using the data processing pipeline with simulated data stream using the expected electronic noise and Lindhard quenching model. In addition of electronic (blue) and nuclear recoils (red), we choose to add the \textit{heat only} (black) background as well. The coloured regions in the quenching plot are defines as the $\pm$ one-$\sigma$ band of the respective event type (NR,ER,HO).}
    \label{fig:Quenching}
\end{figure}

Finally, the detailed simulation confirms the particle identification capabilities of the CryoCube detector, where we confirm a ER-NR discrimination threshold around 100 eV recoil energy as it is shown on Fig.~\ref{fig:Quenching}. This threshold is defined as the overlapping point of the one-$\sigma$ bands for ER and NR in the quenching vs recoil energy plane.  Adding the HO event population shows that they will affect our ability to observe the CE$\nu$NS signal at the lowest energies, and a mitigation of such events is therefore our highest priority.

\subsection{Application to experimental data}
This processing pipeline was also used on experimental data in various situation and with several acquisition devices with great performance. This reflect the high flexibility of this Python-based processing framework which is already being used with various cryogenic detector technologies, namely with NTD-Ge (\cite{hugues, thomas}) and Kinetic Inductance Detectors (\cite{wifikid}).

\section{\textbf{Conclusion}}
We presented an efficient simulation and data processing pipeline dedicated to the CryoCube detector array. Using realistic ionization and heat models (pulse and noise) we confirmed the expected performance of the CryoCube detector array required to observe with high precision the CE$\nu$NS signal from the ILL reactor.  More detailed simulations, including the ILL background and CE$\nu$NS signal, are ongoing to estimate the future \Ricochet{}  CE$\nu$NS precision measurement and new Physics discovery potential. \\

\section*{Acknowledgements}
This work is supported by the European Research Council  (ERC)  under  the  European  Union’s  Horizon  2020  research  and  innovation  program under Grant Agreement ERC-StG-CENNS 803079. We are grateful to J.-B. Filippini and H. Lattaud for their feedbacks on the use of this processing pipeline.

\pagebreak



\end{document}